\UseRawInputEncoding

\documentclass[%
 reprint,
 amsmath,amssymb,
 aps,
floatfix,
]{revtex4-2}
\usepackage{comment}
\usepackage{lipsum}
\usepackage{graphicx}
\usepackage{dcolumn}
\usepackage{bm}
\usepackage{soul, color}
\soulregister\cite7
\soulregister\ref7
\soulregister\pageref7
\usepackage{siunitx}
\usepackage{pxfonts}
\usepackage{chemformula}
\usepackage{float}
\DeclareSIUnit\angstrom{\text {\AA}}
\usepackage[utf8]{inputenc}
\usepackage[T1]{fontenc}
\usepackage{hyperref}

\sethlcolor{white}

\begin{document}

\title{Machine-Learning Force Fields Reveal Shallow Electronic States\\on Dynamic Halide Perovskite Surfaces}

\author{Frederico P. Delgado$^{1}$}
\author{Frederico Simões$^{1}$}%
\author{Leeor Kronik$^{2}$}%
\author{Waldemar Kaiser$^{1}$}%
\author{David A. Egger$^{1,3,4}$}%
\email{david.egger@tum.de}

 \affiliation{%
$^1$Physics Department, TUM School of Natural Sciences, Technical University of Munich, Germany
}%

\affiliation{%
$^2$Department of Molecular Chemistry and Materials Science, Weizmann Institute of Science, Israel
}%

\affiliation{%
$^3$Atomistic Modeling Center, Munich Data Science Institute, Technical University of Munich, Germany
}%

\affiliation{%
$^4$Munich Center for Machine Learning, Munich, Germany
}%

\date{\today}
\begin{abstract}
The spectacular performance of halide perovskites in optoelectronic devices is rooted in their tolerance to defects.
Previous studies showed that defects in these materials generate shallow electronic states.
However, how these shallow states persist amid the pronounced atomic dynamics on halide perovskite surfaces remains unknown.
This work reveals that electronic states at surfaces of prototypical \ch{CsPbBr3} are energetically distributed at room temperature akin to well-passivated inorganic semiconductors, even when covalent bonds remain cleaved and undercoordinated.
Specifically, a striking tendency for shallow surface states is found with approximately 70\% of surface-state energies appearing within \SI{0.2}{eV} or ${\approx}8k_\text{B}T$ from the valence-band edge.
\hl{While these findings do not rule out occurrence of deep traps per se, they} show that even when surface states appear deeper in the gap, they are not energetically isolated and are less likely to act as traps.
We achieve this result by accelerating first-principles calculations via machine learning and show that the unique atomic dynamics in these materials render the formation of deep electronic states at their surfaces unlikely.
These findings reveal the microscopic mechanism behind the low density of deep states at dynamic halide perovskite surfaces, which is key to their device performance.
\end{abstract}

\maketitle

The electronic structure of semiconductor surfaces critically influences device performance, as the cleaving of covalent bonding networks introduces unsaturated or dangling bonds and leads to the formation of surface states~\cite{Heine_1966,basic_theory_of_surface_states, Groß_2003}.
Some of these states are shallow, and can contribute to conductivity or charge transfer. 
Others possess energy levels deep within the gap, where they can act as traps at interfaces in devices and degrade performance.
For example, such traps enhance interface recombination in solar cells, which reduces carrier lifetimes and open circuit voltages.
This has led to extensive research into finding appropriate passivation methods for various types of semiconductors, including traditional ones such as Si and GaAs~\cite{Queisser_Haller_1998}.

Interestingly, \hl{certain experiments point to an} apparent lack of evidence for deep electronic states at non-passivated surfaces of halide perovskite (HaP) semiconductors.
For instance, the Fermi level can be moved in energy throughout much of the band gap when HaPs are deposited on different substrates with varying work functions~\cite{Schulz_Whittaker-Brooks_MacLeod_Olson_Loo_Kahn_2015, Schulz_Tiepelt_Christians_Levine_Edri_Sanehira_Hodes_Cahen_Kahn_2016, Endres_Kulbak_Zhao_Rand_Cahen_Hodes_Kahn_2017, Zohar_Kulbak_Levine_Hodes_Kahn_Cahen_2019,Noel_Habisreutinger_Wenger_Lin_Zhang_Patel_Kahn_Johnston_Snaith_2020,Hong_Kwon_Choi_Le_Kim_Han_Suh_Kim_Sutter-Fella_Jang_2021,Shin_Zu_Cohen_Yi_Kronik_Koch_2021}.
Furthermore, Kelvin-probe measurements established that band bending is small compared to as-grown classical inorganic semiconductors~\cite{Edri_Kirmayer_Mukhopadhyay_Gartsman_Hodes_Cahen_2014, Edri_Kirmayer_Henning_Mukhopadhyay_Gartsman_Rosenwaks_Hodes_Cahen_2014, GualdronReyes_Yoon_Barea_Agouram_MunozSanJose_Melendez_NinoGomez_MoraSero_2019, Kirchartz_Cahen_2020}.
\hl{These previous findings do not automatically imply that photoluminescence and open-circuit voltage measurements will not be influenced by traps nevertheless, because the experiments are affected by recombination processes.
But whether or not electronic states are shallow or deep plays a key role also in these processes, as recombination rates crucially depend on the energetic position of defect states in the band gap of HaP surfaces and interfaces.}

\begin{figure}[t]
\centering
\includegraphics[width=0.5\textwidth]{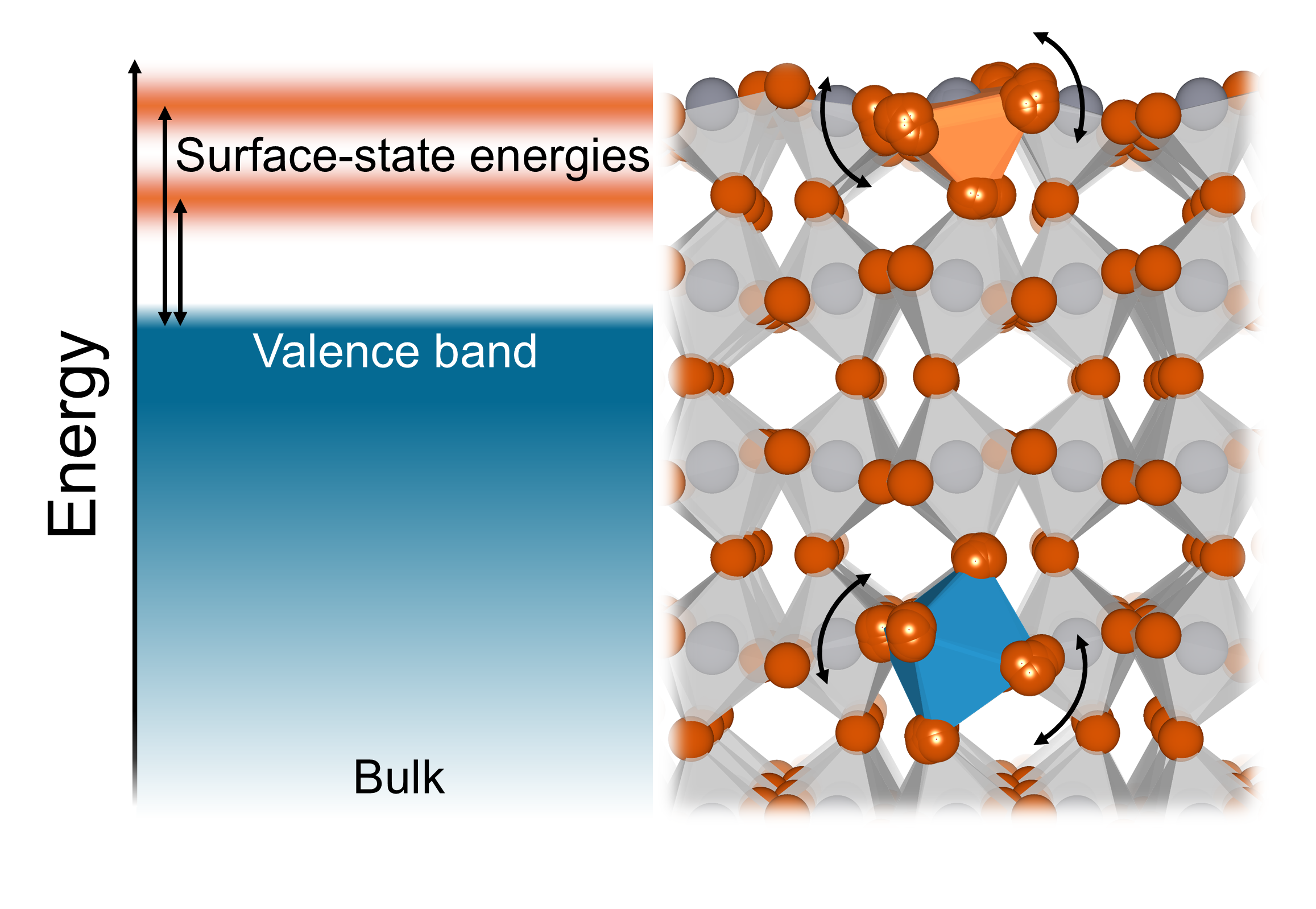}
\caption{Schematic representation of surface-state energies shifting throughout the band gap of a halide perovskite (HaP) semiconductor (left), as a consequence of pronounced atomic dynamics at their surfaces (right).}
\label{fig:1}
\end{figure}
\begin{figure*}
\centering
\includegraphics[width=.7\textwidth]{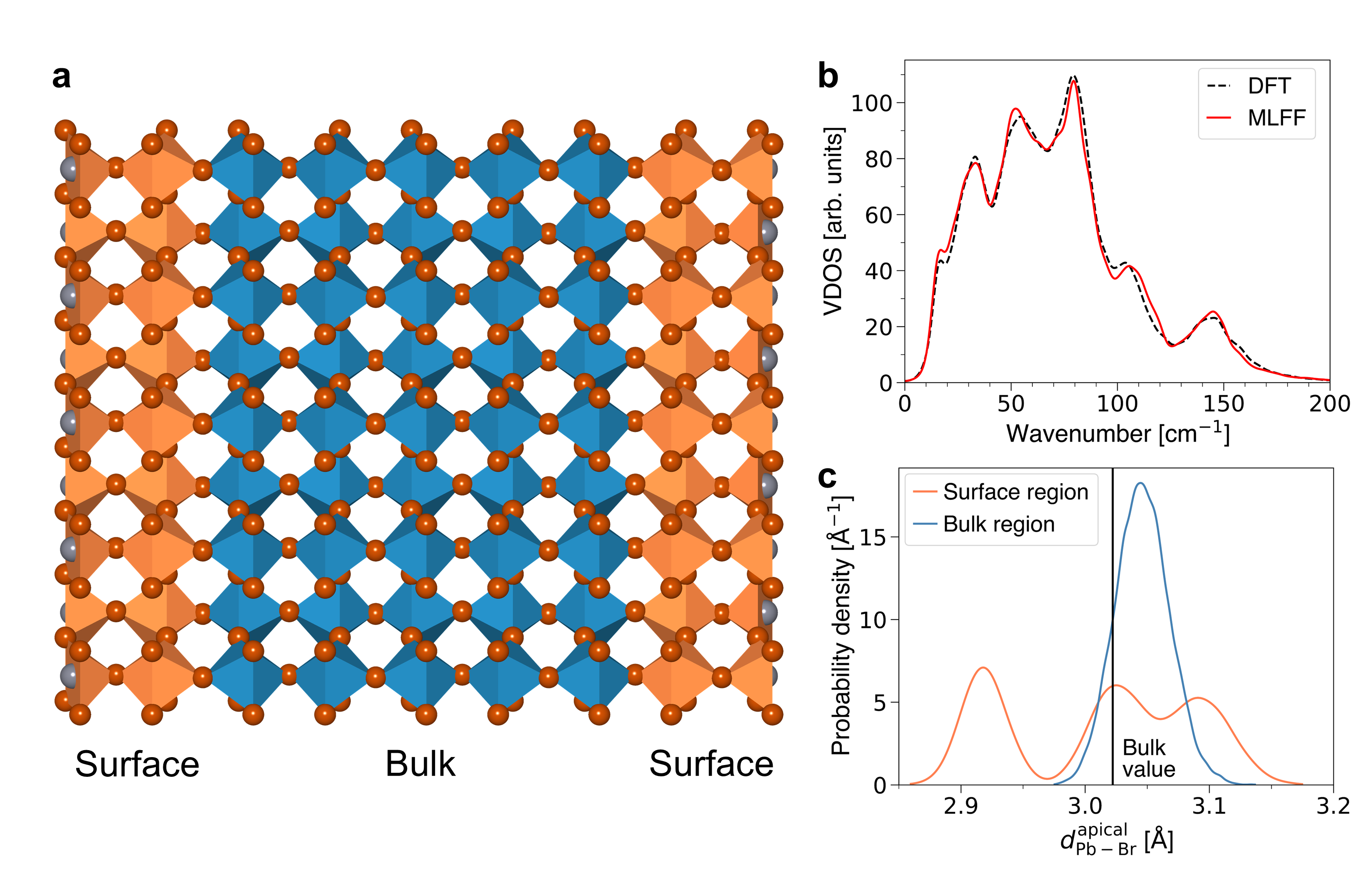}
\caption{\textbf{a.} Structural representation of a $4\times4$ surface supercell of \ch{CsPbBr3}, showing the surface and bulk regions in orange and blue colors, respectively. \textbf{b.} Vibrational density of states (VDOS) of bulk \ch{CsPbBr3} obtained in molecular dynamics (MD) with density functional theory (DFT) and machine-learning force fields (MLFF), highlighting the accuracy of the MLFF. \textbf{c.} Probability density of apical Pb--Br distances along the direction of the surface normal, $d_\mathrm{Pb-Br}^\mathrm{apical}$, identifying dynamic differences in atomic structures across different regions of the slab. Vertical black line indicates the peak of the probability density obtained in a separate calculation of the bulk \ch{CsPbBr3} material.}
\label{fig:2}
\end{figure*}

\hl{When undercoordinated ionic species like lead are present at HaP surfaces, one expects that electronic states in the band gap will arise.
Relevant in this context,} static computational modeling of HaP surfaces indicated that cleaving Pb-halide bonds, responsible for antibonding states at the top of the valence band, stabilizes electronic levels and results in shallow surface states~\cite{database_1_Haruyama2014-ks, database_2_Haruyama2016-zf}.
\hl{The above discussion signifies that the density of deep states at HaP surfaces might be low,} 
despite the presence of at least intrinsic defects from strained and dangling bonds due to terminating the bulk material~\cite{Cahen_Rakita_Egger_Kahn_2024}.
Self-healing effects and defect tolerance in HaPs are actively investigated to rationalize this remarkable observation~\cite{Cohen_Egger_Rappe_Kronik_2019, Rakita_Lubomirsky_Cahen_2019, Ambrosio_Mosconi_Alasmari_Alasmary_Meggiolaro_De_Angelis_2020, Kumar_Hodes_Cahen_2020, Zhang_Turiansky_Van_de_Walle_2020, Cahen_Kronik_Hodes_2021, Ran_Wang_Wu_Liu_MoraPerez_Vasenko_Prezhdo_2023}.

HaPs are unique among ionic-covalent semiconductors with regard to their thermally-triggered atomic motions~\cite{Egger_Rappe_Kronik_2016, Egger_Bera_Cahen_Hodes_Kirchartz_Kronik_Lovrincic_Rappe_Reichman_Yaffe_2018}.
\hl{For instance, these materials show local polar fluctuations}~\cite{Beecher_Semonin_Skelton_Frost_Terban_Zhai_Alatas_Owen_Walsh_Billinge_2016,Yaffe_Guo_Tan_Egger_Hull_Stoumpos_Zheng_Heinz_Kronik_Kanatzidis_etal._2017} \hl{and anharmonic vibrations}\cite{Marronnier_Lee_Geffroy_Even_Bonnassieux_Roma_2017, Marronnier_Roma_Boyer-Richard_Pedesseau_Jancu_Bonnassieux_Katan_Stoumpos_Kanatzidis_Even_2018,
Gehrmann_Egger_2019,Ferreira_Paofai_Létoublon_Ollivier_Raymond_Hehlen_Rufflé_Cordier_Katan_Even_et_al._2020,Klarbring_Hellman_Abrikosov_Simak_2020,
Lanigan-Atkins_He_Krogstad_Pajerowski_Abernathy_Xu_Xu_Chung_Kanatzidis_Rosenkranz_et_al._2021,Lahnsteiner_Bokdam_2022,
Tadano_Saidi_2022,
Fransson_Rosander_Eriksson_Rahm_Tadano_Erhart_2023,Nonlinear_terahertz_2023,
Baldwin_Liang_Klarbring_Dubajic_Dell’Angelo_Sutton_Caddeo_Stranks_Mattoni_Walsh_et_al._2024,Caicedo-Dávila_Cohen_Motti_Isobe_McCall_Grumet_Kovalenko_Yaffe_Herz_Fabini_et_al._2024}.
Strongly anharmonic vibrations were found to critically impact key bulk properties of HaPs, such as carrier mobilities and band gaps~\cite{
Quarti_Mosconi_Ball_D’Innocenzo_Tao_Pathak_Snaith_Petrozza_Angelis_2016,
Gehrmann_Egger_2019,
Schilcher_Robinson_Abramovitch_Tan_Rappe_Reichman_Egger_2021,
Park_Limmer_2023,
Schilcher_Abramovitch_Mayers_Tan_Reichman_Egger_2023,
Seidl_Zhu_Reuveni_Aharon_Gehrmann_Caicedo-Dávila_Yaffe_Egger_2023,
Zacharias_Volonakis_Giustino_Even_2023, Hylton-Farrington_Remsing_2024, Dörflinger_Rieder_Dyakonov, Zhu_Egger_2025}.
Some of the present authors found that atomic dynamics lead to strong defect level fluctuations that could enhance defect tolerance in bulk \ch{CsPbBr3}~\cite{Cohen_Egger_Rappe_Kronik_2019}, with device implications later confirmed by non-adiabatic molecular dynamics~\cite{Wang_Chu_Wu_Casanova_Saidi_Prezhdo_2022}.
At HaP surfaces, where the bonding network is cleaved, atomic motions are expected to be even more pronounced than in the bulk.
A first-principles study on HaP surfaces and nanoparticles highlighted key differences from bulk behavior, notably that formation of vacancies is energetically less favorable at HaP surfaces~\cite{tenBrinck_Zaccaria_Infante_2019}.
Atomic motions create disordered potential energy landscapes that can both localize surface states and shift their energies throughout the band gap (see Fig.~\ref{fig:1}).
Nonetheless, \hl{it is currently debated whether} deep electronic states at HaP surfaces are exceptionally rare~\cite{Cahen_Rakita_Egger_Kahn_2024}.
In addition, a first-principles study by Lodeiro et al.~\cite{database_48_Lodeiro2020-ok} found that atomic dynamics at HaP surfaces do not generate deep states. 
Although unexpected, the mechanistic origin of shallow surface states at dynamic HaP surfaces has received little attention, despite its critical importance for device performance.

Here, we investigate how atomic dynamics at HaP surfaces impact their electronic structure, offering atomic-scale insights into mechanisms that define their functionality in devices. 
We overcome significant bottlenecks associated with computational modeling of thermally-stimulated atomic dynamics in large structures through machine-learning (ML) methods.
Despite pronounced atomic fluctuations at HaP surfaces, we observe a striking tendency for shallow electronic surface states.
Our first-principles calculations resolve this conundrum, revealing that the distinctive atomic dynamics of HaP surfaces make the formation of deep electronic states unlikely. 
These findings elucidate the \hl{low likelihood} of deep electronic states at clean HaP surfaces under ambient conditions, highlighting their advantageous role in semiconductor devices.

\begin{figure*}[t]
\centering
\includegraphics[width=0.7\textwidth]{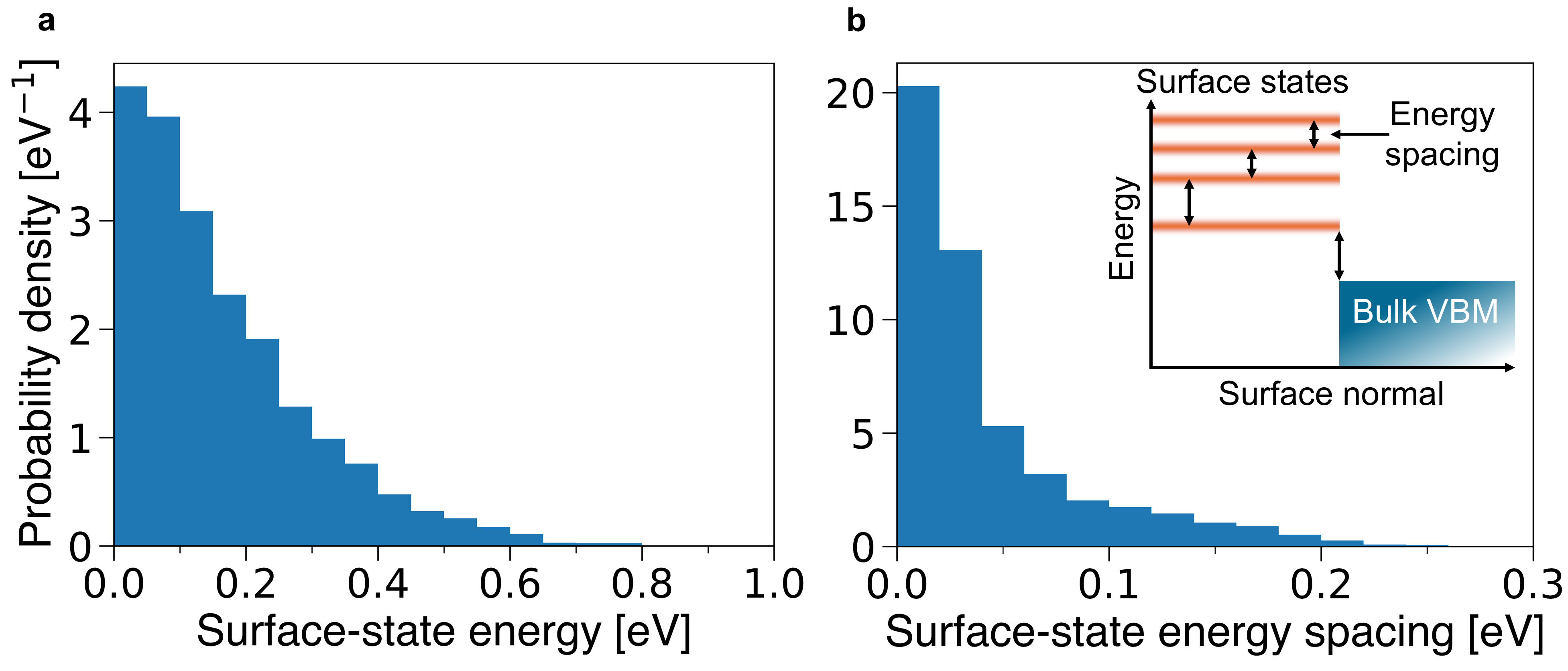}
\caption{\textbf{a.} Probability density of surface-state energies occurring on the surface of \ch{CsPbBr3} at $T=$~\SI{298}{\kelvin}, as calculated with a MLFF and DFT. Zero on the $x$-axis corresponds to the highest occupied state in the bulk region, determined separately for each sampled structure. \textbf{b.} Probability density of the energy spacing between the electronic states on the surface of \ch{CsPbBr3} at $T=$~\SI{298}{\kelvin}. The inset illustrates the energy spacing of surface states.}
\label{fig:3}
\end{figure*}

We focus on atomic dynamics and electronic structure at the surface of \ch{CsPbBr3}. 
\hl{A prototypical HaP material, this system is fully inorganic yet shows similarly pronounced atomic dynamics as related hybrid organic-inorganic variants such as} \ch{MAPbBr3}~\cite{Yaffe_Guo_Tan_Egger_Hull_Stoumpos_Zheng_Heinz_Kronik_Kanatzidis_etal._2017}.
Our computational model employs the repeated-slab approach~\cite{Neugebauer_Scheffler_1992, Srivastava_1997, Zojer_2024}, where dipole corrections are self-consistently calculated in the vacuum region of a 2D-extended surface. 
{We find that} a large structural model is required to reach a desired level of accuracy for surface electronic properties (see Supplementary Information, SI).
To meet this requirement, we employ a structure, consisting of seven full octahedral layers and two cleaved ones in the direction along the surface normal, and laterally expand it in a $4\times4$ surface supercell (see Fig.~\ref{fig:2}a), resulting in a total of {1376} atoms.
This structure allows us to define surface and bulk regions within the slab (see Fig.~\ref{fig:2}a) to characterize the atomic dynamics and their influence on the electronic surface states. 

We perform molecular dynamics (MD) calculations using machine-learning force fields (MLFFs) trained with density functional theory (DFT), see Methods section for details.
MLFFs have become increasingly important for materials modeling in general~\cite{Batatia_Kovács_Simm_Ortner_Csányi_2023, Batzner_Musaelian_Sun_Geiger_Mailoa_Kornbluth_Molinari_Smidt_Kozinsky_2022, Chmiela_Sauceda_Müller_Tkatchenko_2018, Deringer_Bartók_Bernstein_Wilkins_Ceriotti_Csányi_2021, Unke_Chmiela_Sauceda_Gastegger_Poltavsky_Schütt_Tkatchenko_Müller_2021}, and for HaPs in particular~\cite{mlff_2, Bokdam_Lahnsteiner_Sarma_2021, Lahnsteiner_Bokdam_2022, 
Sauceda_Gálvez-González_Chmiela_Paz-Borbón_Müller_Tkatchenko_2022,
Biega_Bokdam_Herrmann_Mohanraj_Skrybeck_Thelakkat_Retsch_Leppert_2023, Liang_Klarbring_Baldwin_Li_Csányi_Walsh_2023,  
Baldwin_Liang_Klarbring_Dubajic_Dell’Angelo_Sutton_Caddeo_Stranks_Mattoni_Walsh_et_al._2024, Pols_Brouwers_Calero_Tao_2023}, as they dramatically lower computational costs and facilitate the study of larger structures through MD simulations.
We observe that accurate predictions of atomic dynamics on HaP surfaces demand a careful training strategy in the ML approach. 
First, we train a MLFF for bulk \ch{CsPbBr3} (see Methods) and find very good agreement for dynamic structural observables such as the vibrational density of states (VDOS, see Fig.~\ref{fig:2}b).
Predicted forces for bulk \ch{CsPbBr3} reach exceptional accuracy compared to DFT and deviate with a root mean square (RMS) error of only \SI{14}{\milli\electronvolt/\angstrom} from DFT forces.
However, when the MLFF trained for the bulk is used for the surface region of \ch{CsPbBr3}, it results in inaccurate predictions and a RMS force error of  \SI{98}{\milli\electronvolt/\angstrom}.
Therefore, we adopt a ML workflow in which we successively train MLFFs specifically for the surface (see Methods section).
This strategy provides us with a MLFF that reaches a RMS force error of \SI{39}{\milli\electronvolt/\angstrom} for the bulk and surface region of the slab.
\hl{While further improvements could be achieved with, e.g., using an $NPT$ ensemble or higher temperatures during training, we consider the present MLFF} satisfactory because it matches the accuracy established in previous benchmark studies of MLFFs in HaPs~\cite{mlff_2, Lahnsteiner_Bokdam_2022, Liang_Klarbring_Baldwin_Li_Csányi_Walsh_2023,Baldwin_Liang_Klarbring_Dubajic_Dell’Angelo_Sutton_Caddeo_Stranks_Mattoni_Walsh_et_al._2024, Pols_Brouwers_Calero_Tao_2023,Oz_2025} and other structurally complex materials~\cite{Miyagawa_Krishnan_Grumet_Baecker_Kaiser_Egger_2024, Wieser_Zojer_2024}.

Fig.~\ref{fig:2}c reports probability densities of apical Pb-Br bond distances along the direction of the surface normal, $d_\mathrm{Pb-Br}^\mathrm{apical}$, recorded in 8333 snapshots across \SI{50}{ps} of a MLFF run at $T=\SI{298}{\kelvin}$.
The distribution for the bulk region in the slab peaks close to the maximum value of bulk \ch{CsPbBr3} at \SI{3.02}{\angstrom}, confirming the validity of our approach. 
A previous study reported that the atomic dynamics on surfaces of the related \ch{MAPbI3} material differ from that of the bulk~\cite{database_48_Lodeiro2020-ok}.
Leveraging ML-accelerated simulations with larger cells, our work provides statistically robust atomic-scale details of surface dynamics.
For the surface region, we observe a density profile that deviates significantly from that of the bulk region, exhibiting three distinct features: 
the peak at \SI{2.92}{\angstrom} corresponds to compressed Pb--Br bonds in the outermost part of the surface, oriented toward the vacuum. 
In contrast, the peak at \SI{3.09}{\angstrom} reflects extended Pb--Br bonds just beneath this outermost layer, marking the transition into the bulk. 
Finally, the third feature, peaking at \SI{3.03}{\angstrom}, represents Pb--Br bonds in the subsequent layer, where the bond lengths have already relaxed to the bulk value.
While these results demonstrate that atomic dynamics in the surface region differ significantly from those in the bulk, they also show that dynamic surface-induced disruptions quickly diminish within less than \SI{1}{\nm} away from the surface into the material.

\begin{figure*}
\centering
\includegraphics[width=.7\textwidth]{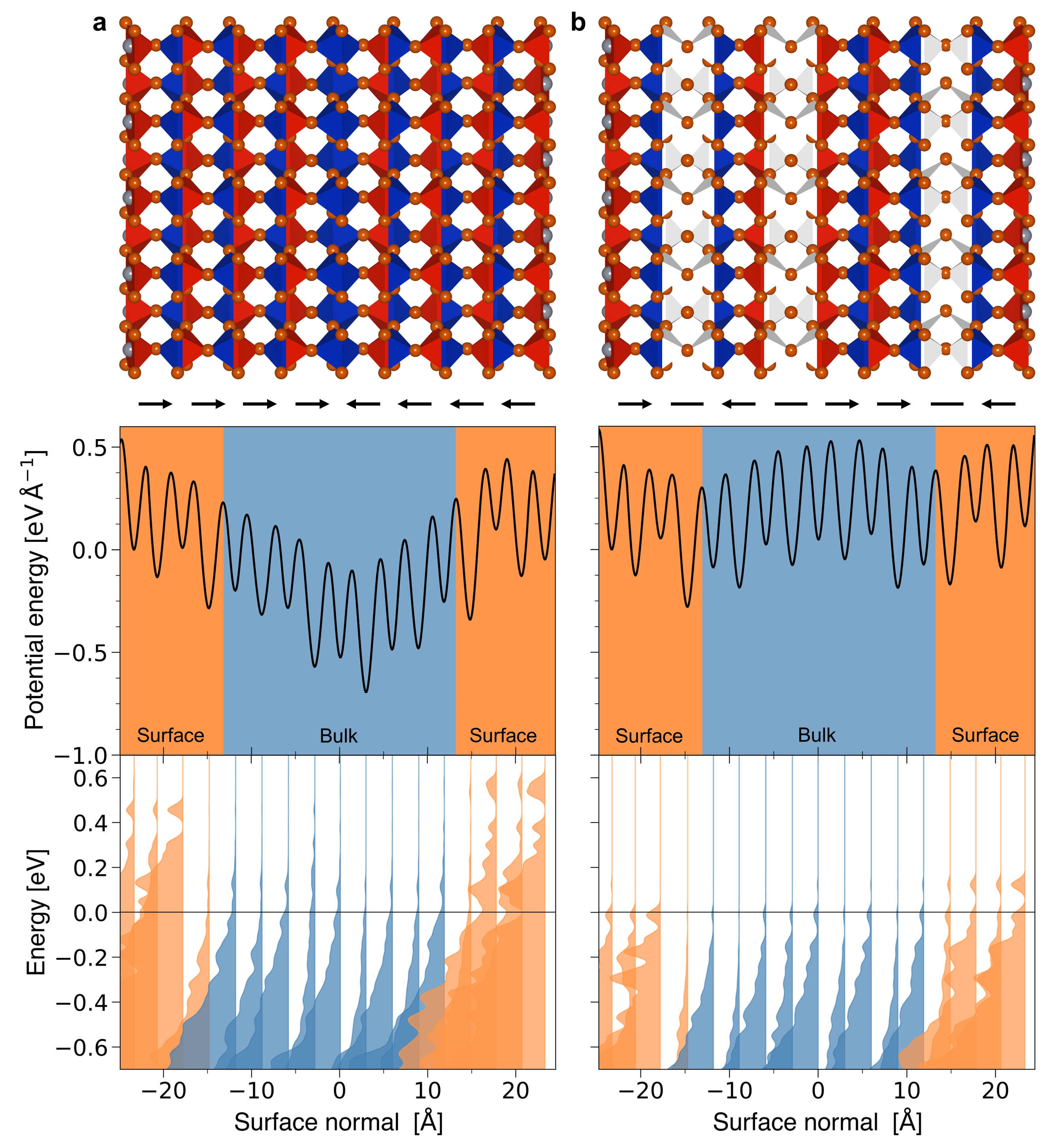}
\caption{\hl{Schematic representations of snapshots of atomic dynamics and} associated dipole moments, and their effect on a moving-average of the plane-averaged potential energy profile \hl{and layer-resolved density of states} (see Methods section) across the slab. \textbf{a.} When successive compression and extension in adjacent apical Pb-Br bonds across the slab occur (top panel), they create a series of dipole moments (black arrows) and surface regions that are significantly higher in energy than the bulk region, resulting in deep surface states. Closely-spaced electronic states fill the energy gap between deeper surface states and the bulk electronic structure, as seen in the layer-projected density of states (bottom panel). \textbf{b.} By contrast, absence of spatially coordinated compression and extension, and lack of spatially coordinated dipole moments, results in a constant potential profile through the slab and shallow surface states. Zero on the $x$-axis corresponds to the center of the slab and the vacuum region was removed from the potential energy analysis to prevent boundary effects from influencing the smoothing of the potential energy profile. The layer-projected density of states (bottom panel) shows an electronic structure that is energetically more aligned between bulk and surface region. Zero on the $y$-axis of the bottom panels refers to the highest occupied state in the bulk region, indicated by a thin horizontal line.}
\label{fig:4}
\end{figure*}

In light of the significant surface-induced structural perturbations, we now investigate their influence on electronic surface states in \ch{CsPbBr3}.
We sampled 312 structures from the equilibrated MLFF trajectory at equidistant intervals of \SI{96}{fs}, allowing a statistical exploration of the electronic structure at $T=\SI{298}{\kelvin}$ through self-consistent DFT calculations.
\hl{We emphasize the importance of following a statistical approach: a focus on average electronic quantities can miss effects due to tails which can dominate recombination processes, and a focus on average structures misses the effect of thermally-stimulated atomic motions. 
Our analysis is rooted in determining the localization of electronic states and identifies surface states as being predominantly localized in the surface region of the slab (see Methods).}
We find a large number of surface states in many of the extracted snapshots, which can contain up to several dozens of electronic states concentrated in the surface region of the $4\times4$ supercell we employ (see SI).
The specific shape of the calculated probability density is reminiscent of a ``U-shaped'' distribution of surface states that are predominantly shallow (see Fig.~\ref{fig:3}a)~\cite{Flietner_1974}. 
This is remarkable because such a shape is indicative of well-passivated semiconductor surfaces~\cite{Allen_Gobeli_1962, Yamagishi_1968, Lam_1973,Flietner_1974, Angermann_Dittrich_Flietner_1994, Kronik_Shapira_1999}, yet strained and dangling covalent bonds in \ch{CsPbBr3} were explicitly present in our calculations and no passivation was applied.
Specifically, almost 70\% of all recorded surface-state energies are within \SI{0.2}{eV} or ${\approx}8k_\text{B}T$ from the valence-band edge. 

{For further insights,} we calculate the energy spacing between all surface states in \ch{CsPbBr3} (see Fig.~\ref{fig:3}b). 
This is important because when electronic states appear deeper in the band gap, they only act as charge traps when they are energetically isolated.
We find that this is not the case for \ch{CsPbBr3}, because the calculated energy spacings between consecutive surface states are extremely small, with approximately 90\% of surface states separated by less than \SI{0.1}{eV}.
\hl{In passing, we note that preliminary test calculations using a hybrid functional as well as spin-orbit coupling show that while the depth of surface states remains similar, the energy spacing between them could be more sensitive to the choice of DFT functional (see SI).}
Altogether, despite significant structural fluctuations on the surface, the electronic states remain largely shallow and closely spaced in energy. 
\hl{These findings do not rule out the appearance of deeper states as such, and in fact our calculated distribution of surface-state energies finds a tail deeper into the gap} (see Fig.~\ref{fig:3}a). 
\hl{Rather, our results imply} that even when surface states appear deeper in the gap, carriers can quickly thermalize to the band edges, which prevents trapping. 
These findings indicate a degree of resilience in the electronic structure of \ch{CsPbBr3} against surface-induced perturbations.

We now turn to an analysis of the combined atomic, electrostatic, and electronic dynamics on \ch{CsPbBr3} surfaces at $T=\SI{298}{\kelvin}$.
We select representative snapshots from 312 DFT calculations performed on the MLFF trajectory, each capturing surface states appearing to be either deep or shallow in the band gap (see Fig.~\ref{fig:4}a and b).
In case of a state that appears deeper in the band gap, we find a repeating pattern where $d_\mathrm{Pb-Br}^\mathrm{apical}$, averaged per layer, switches between compression and elongation (cf. Fig.~\ref{fig:2}c).
Notably, this introduces a characteristic series of local dipoles per layer across the slab (see Fig.~\ref{fig:4}a, top panel). 
The collective effect of these dipoles substantially alters the electrostatic potential~\cite{Natan_Kronik_Haick_Tung_2007, Zojer_2024}, elevating the energy of the surface regions well above that of the bulk (Fig.~\ref{fig:4}a, middle panel).
As a consequence, electronic states in the surface region are shifted to higher energies, extending further into the band gap and giving rise to states that appear deeper in the band gap (Fig.~\ref{fig:4}a, bottom panel).
{However}, even when a surface state appears deeper in the band gap, the layer-projected density of states \hl{-- which resolves spatial localization of electronic states along the surface normal --} exhibits an almost continuous density of electronic states emerging with it at the surface. 
\hl{This finding} is in line with the minute energy spacings of surface states reported above (cf. Fig.~\ref{fig:3}b). 
The closely-spaced surface states essentially fill the energy gap between deep states and the bulk electronic structure.
Notably, our results show that new electronic states are not generated in this scenario.
Rather, the electronic structure at the surface of \ch{CsPbBr3} is shifted upwards with the potential energy as a direct consequence of the collective electrostatic effect of aligned dipoles, which does not isolate surface states energetically.
By contrast, we find that an absence of a repeating pattern of $d_\mathrm{Pb-Br}^\mathrm{apical}$ prevents a characteristic sequence of dipoles across the slab when surface states are shallow (see Fig.~\ref{fig:4}b, top panel).
Consequently, the surface and bulk region are energetically much more aligned (Fig.~\ref{fig:4}b, middle panel), allowing electronic states at the edges of the slab to remain shallow (bottom panel).

\begin{figure}
\centering
\includegraphics[width=0.4\textwidth]{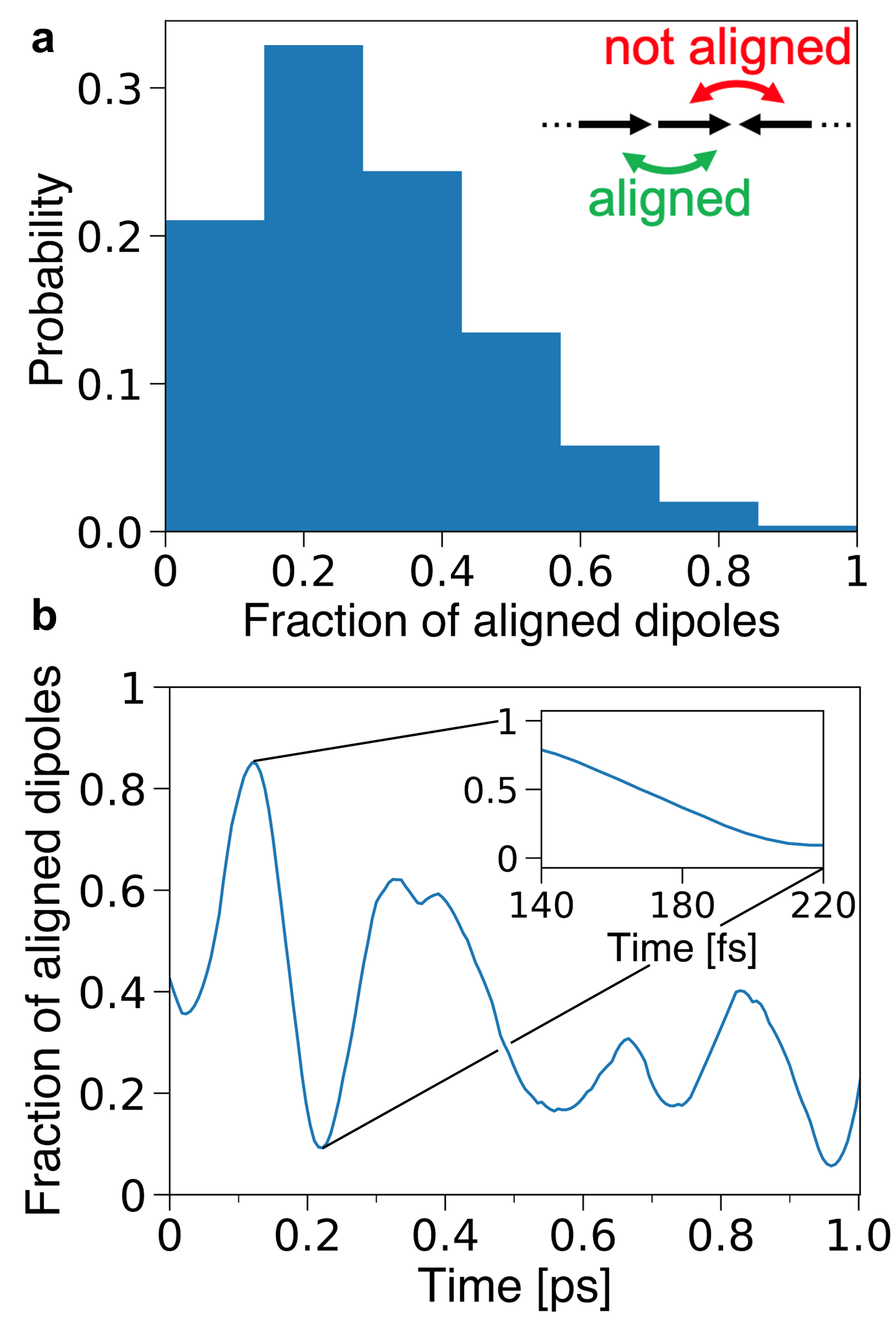}
\caption{\textbf{a.} Probability of finding adjacent polar layers with aligned dipole moments in a \ch{CsPbBr3} slab. \hl{We determine the orientation of the layer dipole moment using a structural proxy, i.e.,} from patterns in $d_\mathrm{Pb-Br}^\mathrm{apical}$ across a \SI{50}{ps} MLFF run. The inset illustrates examples of aligned and non-aligned layer dipole moments. \textbf{b.} Time evolution of the fraction of adjacent polar layers with aligned dipole moments over \SI{1}{ps} of the MLFF simulation, obtained by smoothing the data from panel a for easy visualization. The inset highlights the rapid relaxation of a higher fraction within tens of femtoseconds.}
\label{fig:5}
\end{figure}

Importantly, the occurrence of repeating dipolar patterns is thermodynamically far less likely than a scenario where dipoles are disoriented.
{Atomic dynamics in the soft lattices of HaPs and their surfaces are accompanied by relatively small energy changes. 
Entropic contributions are sizable in comparison because HaPs are strongly dynamically disordered already around room temperature}~\cite{Wu_Tan_Shen_Hu_Miyata_Trinh_Li_Coffee_Liu_Egger_et_al._2017, Gehrmann_Egger_2019, Weadock_Sterling_Vigil_Gold-Parker_Smith_Ahammed_Krogstad_Ye_Voneshen_Gehring_et_al._2023}.
{To investigate the interplay between enthalpy and entropy,} we use \hl{a structural proxy: we compute $d_\mathrm{Pb-Br}^\mathrm{apical}$ to determine} the dipole orientation per layer across the \ch{CsPbBr3} slab (see Methods section). 
Inspecting \SI{50}{ps} of the MLFF run, we find that a series of aligned dipoles rarely occurs and that they are much more likely to be disoriented (see Fig.~\ref{fig:5}a). 
Furthermore, we find that once a series of aligned polar layers spontaneously emerges, it relaxes quickly within tens of femtoseconds (see Fig.~\ref{fig:5}b), ensuring it does not persist long enough to act as a trap.
{As a result, entropic effects prevent the surface from developing a series of polar layers and short-ranged correlations in the }bulk\cite{Wu_Tan_Shen_Hu_Miyata_Trinh_Li_Coffee_Liu_Egger_et_al._2017, Gehrmann_Egger_2019, Weadock_Sterling_Vigil_Gold-Parker_Smith_Ahammed_Krogstad_Ye_Voneshen_Gehring_et_al._2023} are maintained at surfaces of HaPs.
Thus, the prevalence of disordered dipoles over repeating patterns reinforces the tendency toward energetically aligned surface and bulk regions and shallow surface states in \ch{CsPbBr3}.

Our findings indicate that the combination of defect-tolerant electronic structure and unique atomic dynamics in HaPs synergistically minimizes the occurrence of deep states on their surfaces.
In traditional inorganic semiconductors, intrinsic defects at surfaces due to strained or dangling covalent bonds create deep surface states.
For these materials, a ``U-shaped'' distribution with a low density of surface states emerges once their surfaces are passivated, for instance via hydrogenation or oxidation in case of Si~\cite{Allen_Gobeli_1962, Yamagishi_1968, Lam_1973,Flietner_1974, Angermann_Dittrich_Flietner_1994, Kronik_Shapira_1999}.
Here, we find this characteristic distribution of surface states for non-passivated surfaces of \ch{CsPbBr3} where all bonds are undercoordinated.
\hl{Notably, we report that roughly 70\% of surface-state energies appear within 0.2\,eV from the edge of the valence band.
This is in agreement with numerical simulations that found neutral, shallow defect states located only 150\,meV from the valence-band edge to best explain transient photoluminescence data of layer stacks including MAPbI$_\mathrm{3}$ films~\cite{Krückemeier_2021}.
While our findings do not preclude the potential benefits of passivation for HaP surfaces, they do indicate that surface passivation may be of secondary importance given that even non-passivated surfaces show beneficial electronic properties.}

Previous studies have suggested that semiconductors with antibonding states at the top of the valence band are more likely to exhibit tolerance to bond cleavage and can accommodate defects better ~\cite{zakutayev2014defect,database_73_Brandt_2017,database_76_Kovalenko_2017}.
Static DFT calculations showed that the antibonding character of the valence band edge in HaPs leads to shallow states, both in case of point defects~\cite{Yin_Shi_Yan_2014} and surfaces~\cite{database_1_Haruyama2014-ks, database_2_Haruyama2016-zf}.
In line with our findings, a previous study using dynamic calculations~\cite{database_48_Lodeiro2020-ok} found shallow states at HaP surfaces.
However, the atomic dynamics at HaP surfaces are pronounced and how this can be reconciled with a low density of surface states remained unknown, especially given that such dynamics can cause the appearance of deep levels in the bulk~\cite{Cohen_Egger_Rappe_Kronik_2019, Wang_Chu_Wu_Casanova_Saidi_Prezhdo_2022}.
Our study establishes that in case of HaP surfaces the atomic dynamics suppress the \hl{likelihood of} formation of deep states and thereby advantageously support their outstanding performance in semiconductor devices.
\hl{Given that vibrational properties of HaPs depend on the ionic composition, we expect that the density of deep surface states will depend on it too and that, e.g., presence of an organic A-site cation could influence trapping behavior.}

In summary, we studied the effect of pronounced atomic dynamics at HaP surfaces on the occurrence of surface states due to strained and dangling bonds. 
By employing MLFFs to accelerate dynamic calculations of \ch{CsPbBr3} surfaces, we observe an electronic state distribution akin to well-passivated semiconductor surfaces, despite the presence of undercoordinated covalent bonds at the bulk termination.
Our first-principles calculations resolve the puzzling observation that HaPs lack a high density of surface states despite pronounced atomic dynamics. 
We find that the soft lattice and significant atomic fluctuations suppress the formation of deep surface states and beneficially impact defect tolerance in HaPs.
Specifically, surface states appearing deeper in the gap are rare, they vanish on the order of tens of femtoseconds, and even when they are present they are not energetically isolated, all of which prevent formation of deep traps for electronic charge carriers.
\hl{While these findings do not rule out the formation of deep traps at HaP surfaces as such, they do imply that their occurrence is rare in comparison to non-passivated surfaces of traditional inorganic semiconductors.}
Our study provides important microscopic insights on the surprising absence of deep electronic states at clean HaP surfaces, which is critical for the performance of these materials in semiconductor devices.

\section*{\label{sec:methods}Computational Methods}
\textbf{Density functional theory:} DFT calculations were performed within the Vienna Ab Initio Simulation Package (VASP)~\cite{PhysRevB.54.11169_1996, hafner2008ab}.
The outermost \textit{s}, \textit{p}, and \textit{d} (in case of Pb) electrons were treated as valence electrons and core-valence electron interactions described using the projector-augmented wave (PAW) method~\cite{PhysRevB.59.1758}.
The Perdew-Burke-Ernzerhof (PBE)~\cite{PhysRevLett.77.3865} exchange-correlation functional was used, and dispersion corrections were accounted for using the Tkatchenko-Scheffler method~\cite{PhysRevLett.102.073005} with iterative Hirshfeld partitioning~\cite{Bultinck_Van_Alsenoy_Ayers_Carbó-Dorca_2007}. 
In all calculations, a plane-wave energy cutoff of \SI{250}{\eV} was employed. 
The Brillouin zone of the bulk supercell was sampled with a $3\times3\times2$ $\Gamma$-centered \textit{k}-point grid.

\textbf{Surface geometry:} We first fully relaxed (cell shape, volume, and ion positions) a $2\times2\times2$ supercell of \ch{CsPbBr3} in its orthorhombic phase. 
The resulting cell parameters are $a= \SI{17.19}{\angstrom} $, $b=\SI{15.84}{\angstrom}$, $c=\SI{23.48}{\angstrom}$ and $\alpha=\beta=\gamma={90^\circ}$.
Starting from the optimized bulk structures, we created \ch{CsPbBr3} surface slabs with PbBr$_2$ termination along the (001) direction. 
The thickness of the surface slab was chosen such that the normalized electronic density of states (DOS) was well converged (see SI). 
The final supercell model consists of 7 complete octahedral layers, which are terminated in both directions by a PbBr$_2$ surface layer that contains under-coordinated Pb ions.
In addition, a vacuum layer of \SI{15}{\angstrom} was added.
In all surface calculations, dipole corrections were applied along the $z$ direction to avoid spurious interactions between periodic images.
We created two surface models, one with $2\times2$ and the final one with $4\times4$ supercells along $x$ and $y$ direction, containing 344 and 1376 atoms, respectively.
\textit{k}-point sampling was set to $2\times2\times1$ and $\Gamma$-only for the $2\times2$ and the $4\times4$ slab models, respectively.

\textbf{ML workflow:} MLFF runs were performed using the Bayesian regression scheme implemented in VASP 6.4~\cite{mlff_1, mlff_2, mlff_desc}.
MLFFs were trained on-the-fly of \textit{ab initio} molecular dynamics simulations at a temperature of \SI{298}{\kelvin} using the $NVT$ ensemble and the Nos\'e-–Hoover thermostat~\cite{Nosé_1984, Hoover_1985}. 
We chose a time step of \SI{6}{\fs} for the integration of the dynamical equations of motion. 

The ML workflow to create the MLFF used in our study included three successive training runs, with the MLFF resulting from each stage serving as the starting point for the subsequent optimization.
Each training run lasted \SI{100}{\ps}, ensuring the robustness and convergence of the final MLFF (see SI for discussion of hyperparameters).
First, an initial MLFF has been generated for the bulk supercell of \ch{CsPbBr3}, where 131 structures have been explicitly computed with DFT. 
Second, a training run for the $2\times2$ surface model was performed, which gathers an additional 162 structures. 
The training is completed with a final training run for the $4\times4$ surface supercell, adding 65 more structures to the total of 358 structures that the final MLFF has explicitly been trained on. 

The final MLFF was used to generate a \SI{100}{\ps} trajectory for the $4\times4$ surface model, with the initial \SI{50}{\ps} being discarded as equilibration time. 
The resulting trajectory provides an ensemble from which explicit snapshots are selected to conduct the statistical analysis of surface states.

\textbf{Analysis:}
\paragraph{Apical bond calculations\\}

Pb--Br apical bond lengths, $d_\mathrm{Pb-Br}^\mathrm{apical}$, were calculated for each Pb--Br pairs that share a bond oriented along the surface normal.
For our statistical analysis, we compute the average across all apical Pb--Br bonds in each layer.

\paragraph{Surface state calculations\\}

We sampled 312 structures from the equilibrated MLFF trajectory with equidistant spacing of \SI{96}{fs}. For each structure, a DFT single-point calculation was performed and the charge densities of the 60 energetically highest Kohn-Sham orbitals were extracted.
These were then categorized into surface states and bulk states, with surface states being identified if \SI{80}{\percent} of the charge density, associated with a given electronic state, resides in the surface region, which is composed of the outermost one-and-a-half octahedra of each surface (see Fig.~2, main text).
States which do not fulfill this criterion are classified as bulk states. 

Occupied bulk states with the highest orbital energy are identified as the bulk valence band maximum for each of the 312 structures, as illustrated in Fig.~\ref{fig:1}.
All surface states with energies above the bulk valence band maximum are taken into account in the statistical analysis of the surface state energies conducted in this work.

{After identifying} all surface states and their energies, the energy spacing between consecutive surface states was calculated for each of the 312 sampled structures, including the difference between the lowest energy surface state and the bulk valence band maximum, as illustrated in the inset of Fig.~\ref{fig:4}b. 
We calculated the energy spacing for each side of the slab separately and then combined the recorded statistics into one probability density.
{Furthermore, we calculated the layer-projected density of states} by projecting the electronic structure on consecutive Pb-Br and Br-Cs layers along the surface normal of the slab, respectively, and normalizing each curve separately.

\paragraph{Potential energy analysis\\}

Local potential energy including the ionic and Hartree contributions were extracted for all 312 considered structures. 
To analyze the local potential energy across the surface and in the bulk regions, we computed planar averages of the potential along the surface normal within VASPKIT~\cite{VASPKIT}.
We calculated a moving-average by smoothing the plane-averaged potential using a convolution with window size equivalent to an octahedron's thickness.
Smooth potential energy profiles were used in all our analysis.

\paragraph{Analysis of dipole moments across the slab\\}

The concerted alignment of dipoles in the slab was assessed throughout the MLFF run. 
To this end, the slab was divided into 8 equal domains along the direction of the surface normal, each containing multiple Pb--Br--Pb bonds in the apical direction. 
The difference between the two apical bond distances, $d_\mathrm{Pb-Br}^\mathrm{apical}$, is calculated for each of the Pb--Br--Pb bonds in each of the domains. 
The average of all differences across each domain was used as a proxy for the dipole moment in each domain, applying a threshold of \SI{0.03}{\angstrom} below which no dipole was assigned. 
The dipoles per layer determined in this way were then used to evaluate the alignment across neighbouring domains. 
The number or neighbouring domain pairs with aligned dipoles was summed up and normalized by the maximum number of possible pairs. 
The resulting fraction of aligned dipoles was monitored throughout time and smoothed with a Savitzky-Golay filter for easy visualization.

\section*{Acknowledgements}
We thank David Cahen (Weizmann Institute of Science), Antoine Kahn (Princeton University), \hl{and Thomas Kirchartz (FZ Jülich)} for fruitful discussions.
Funding provided by Germany's Excellence Strategy – EXC 2089/1-390776260 , and by TUM.solar in the context of the Bavarian Collaborative Research Project Solar Technologies Go Hybrid (SolTech), are gratefully acknowledged. The authors further acknowledge the Gauss Centre for Supercomputing e.V. for funding this project by providing computing time through the John von Neumann Institute for Computing on the GCS Supercomputer JUWELS at Jülich Supercomputing Centre.

\bibliography{main_text}
\end{document}